\documentclass[12pt]{article}
\usepackage{pdproc,epsfig} 
\def\ltap{\ \raisebox{-.4ex}{\rlap{$\sim$}} \raisebox{.4ex}{$<$}\ }
\def\gtap{\ \raisebox{-.4ex}{\rlap{$\sim$}} \raisebox{.4ex}{$>$}\ }
\newcommand{\dms}{\mbox{$\Delta m^2_{\odot}$ }}
\newcommand{\dma}{\mbox{$\Delta m^2_{\rm A}$ }}

\newcommand{\deltaatm}{\mbox{$\Delta  m^2_{\mathrm{atm}} \ $}}
\newcommand{\deltasol}{\mbox{$ \Delta  m^2_{\odot} \ $}}

\newcommand{\betabeta}{\mbox{$(\beta \beta)_{0 \nu}  $}}
\newcommand{\mefff}{\mbox{$ < \! m  \! > $}}
\newcommand{\meff}{\mbox{$\left|  < \! m  \! > \right| \ $}}
\newcommand{\hbeta}{$\mbox{}^3 {\rm H}$ $\beta$-decay \ }
\newcommand{\eV}{\mbox{$ \  \mathrm{eV} \ $}}
\newcommand{\deltatre}{\mbox{$ \ \Delta m^2_{32} \ $}}
\newcommand{\deltadue}{\mbox{$ \ \Delta m^2_{21} \ $}}

\newcommand{\ts}{\mbox{$\tan^2
\theta_\odot$}}

\newcommand{\deltaatmmax}{\mbox{$(\Delta  m^2_{\mathrm{atm}})_{ \! 
\mbox{}_{\mathrm{MAX}}} \ $}}
\newcommand{\deltaatmmin}{\mbox{$(\Delta  m^2_{\mathrm{atm}})_{ 
\! \mbox{}_{\mathrm{MIN}}} \ $}}

\begin{document}
\renewcommand{\thefootnote}{\alph{footnote}}
{\flushright
Ref. SISSA 67/03/EP

UCLA/03/TEP/21

hep-ph/0308034

}

\title{NEUTRINO MASSES, MIXING AND $\betabeta-$ DECAY
}
\author{S. PASCOLI}
\address{Department of Physics and Astronomy,\\ University of California,
Los Angeles CA 90095-1547, USA}

\author{S. T. PETCOV 
\footnote{Invited talk given at the Xth International 
Workshop on Neutrino Telescopes,
March 11 - 14, 2003, Venice, Italy; to be published in the 
Proceedings of the Workshop.}~
\footnote{Also at: INRNE, Bulgarian Academy of Sciences, 1789 Sofia, 
Bulgaria.}
}
\address{Scuola Internazionale Superiore di Studi Avanzati, and\\
         INFN - Sezione di Trieste, I-34014 Trieste, Italy
}
\vspace{-0.5cm}
\abstract{
The predictions for the effective
Majorana mass $\meff$ in $\betabeta-$decay
in the case of 3-$\nu$ mixing 
and massive Majorana neutrinos are 
reviewed. The physics potential of the 
experiments, searching for $\betabeta-$decay 
and having sensitivity to $\meff \gtap 0.01$ eV,
for providing information on the type of 
the neutrino mass spectrum,
on the absolute scale of neutrino masses 
and on the Majorana CP-violation phases in 
the PMNS neutrino mixing matrix is
also discussed.
}
   
\normalsize\baselineskip=15.25pt

\section{Introduction}
\vspace{-0.3cm}

 The solar neutrino experiments
Homestake, Kamiokande, SAGE, GALLEX/GNO,
Super-Kamiokande (SK) and SNO \cite{Cl98,SKsol,SNO1,SNO2}, 
the data on atmospheric neutrinos
obtained by the Super-Kamiokande (SK) 
experiment \cite{SKatm00} and 
the results from the KamLAND 
reactor antineutrino 
experiment \cite{KamLAND}, 
provide very strong 
evidences for oscillations 
of flavour neutrinos
driven by nonzero neutrino masses 
and neutrino mixing. 
The evidences for solar 
$\nu_e$ oscillations into 
active neutrinos $\nu_{\mu,\tau}$,
in particular, 
were spectacularly reinforced 
by the first data from the SNO 
experiment \cite{SNO1}
when combined with the data from the 
SK experiment \cite{SKsol}, 
by the more recent SNO data \cite{SNO2},
and by the first results of the KamLAND 
\cite{KamLAND} experiment.
Under the rather plausible 
assumption of CPT-invariance,
the KamLAND data practically 
establishes \cite{KamLAND} the
large mixing angle (LMA)
MSW solution as unique solution
of the solar neutrino problem.
This remarkable result
brings us, after more than 
30 years of research, 
initiated by the pioneer
works of B. Pontecorvo \cite{Pont4667} and the
experiment of R. Davis et al. \cite{Davis68},
very close to a complete understanding of the 
true cause of the solar neutrino problem. 

  The combined analyses of the available
solar neutrino and KamLAND  data, performed 
within the two-neutrino mixing hypothesis,
identify two distinct solution sub-regions within 
the LMA solution region~
(see, e.g., \cite{fogli,band,bahcall}).
The best fit values of  the two-neutrino 
oscillation parameters - the solar neutrino mixing
angle $\theta_{\odot}$ 
and the mass squared difference
$\deltasol$, in the two
sub-regions - low-LMA or LMA-I, and high-LMA or LMA-II,
are given by (see, e.g., \cite{fogli}):
\begin{eqnarray}
\label{deltasolI}
\deltasol^{I} = 7.3 \times 10^{-5}~{\rm eV^2}, \ \ & \ \ 
\tan^2 \theta_\odot^{I} = 0.46,  \\
\label{deltasolII}
\deltasol^{II} = 1.5 \times 10^{-4}{\rm eV^2}, \ \
& \ \
\tan^2 \theta_\odot^{II} = 0.46.
\end{eqnarray}
%
\noindent The LMA-I solution is preferred 
statistically by the data. At 90\% C.L. 
one finds \cite{fogli}: 
\begin{equation}
\deltasol \cong (5.6 - 17) \times 10^{-5}~{\rm eV^2},~~~ 
\tan^2 \theta_\odot \cong (0.32 - 0.72)~.
\label{sol90}
\end{equation}
%
\indent  The observed Zenith angle dependence of 
the multi-GeV $\mu-$like events in the Super-Kamiokande
experiment unambiguously demonstrated
the disappearance of the atmospheric  
$\nu_{\mu}$ ($\bar{\nu}_{\mu}$) 
on distances $L \gtap 1000$ km.
The Super-Kamiokande (SK) atmospheric neutrino data
is best described
in terms of dominant $\nu_{\mu} \rightarrow \nu_{\tau}$ 
($\bar{\nu}_{\mu}\rightarrow \bar{\nu}_{\tau}$) oscillations
with (almost) maximal mixing and 
neutrino mass squared difference of  
$|\dma| \cong (1.8 - 4.0)\times 10^{-3}$ eV$^2$ 
($90\%$ C.L.) ~\cite{SKatm00}.
According to the more recent 
combined analysis of the data from 
the SK and K2K experiments
\cite{fogliold} one has:
\begin{equation} 
2.1 \,\times\, 10^{-3}\,\mbox{eV}^2\,\ltap\,
|\dma|\, \ltap\,3.3\, \times\,10^{-3}\,\mbox{eV}^2~,
~~~~~90\%~{\rm C.L.}
\label{atmo90}
\end{equation}
%
\indent   The interpretation of the solar and
atmospheric neutrino, and of the KamLAND
data in terms of 
neutrino oscillations requires
the existence of 3-neutrino mixing
in the weak charged lepton current 
(see, e.g., \cite{BGG99}):
\begin{equation}
\nu_{l \mathrm{L}}  = \sum_{j=1}^{3} U_{l j} \, \nu_{j \mathrm{L}}~.
\label{3numix}
\end{equation}
\noindent Here $\nu_{lL}$, $l  = e,\mu,\tau$,
are the three left-handed flavor 
neutrino fields,
$\nu_{j \mathrm{L}}$ is the 
left-handed field of the 
neutrino $\nu_j$ having a mass $m_j$
and $U$ is the Pontecorvo-Maki-Nakagawa-Sakata (PMNS)
neutrino mixing matrix \cite{BPont57}.
The PMNS mixing matrix $U$ can be parametrized by 
three angles, $\theta_{\rm atm}$,
$\theta_{\odot}$, and $\theta$,
and, depending on whether the massive 
neutrinos $\nu_j$ are Dirac or Majorana particles -
by one or three CP-violating phases \cite{BHP80,Doi81}.
In the standard parametrization 
of $U$ (see, e.g., \cite{BGG99}) 
the three mixing angles are denoted as
$\theta_{12}$, $\theta_{13}$ and
$\theta_{23}$:
\begin{equation}
U = \left(\begin{array}{ccc} 
c_{12}c_{13} & s_{12}c_{13} & s_{13}\\
 - s_{12}c_{23} - c_{12}s_{23}s_{13}e^{i\delta} & 
c_{12}c_{23} - s_{12}s_{23}s_{13}e^{i\delta} & s_{23}c_{13}e^{i\delta}\\
s_{12}s_{23} - c_{12}c_{23}s_{13}e^{i\delta} 
& -c_{12}s_{23} - s_{12}c_{23}s_{13}e^{i\delta}
& c_{23}c_{13}e^{i\delta}\\ 
\end{array} \right)
~{\rm diag}(1,e^{i\frac{\alpha_{21}}{2}},e^{i\frac{\alpha_{31}}{2}})
\label{Umix}
\end{equation}
%
\noindent where we have used the usual notations, 
$s_{ij} \equiv \sin \theta_{ij}$,
$c_{ij} \equiv \cos \theta_{ij}$, 
$\delta$ is the Dirac CP-violation phase
and $\alpha_{21}$ and $\alpha_{31}$ 
are two Majorana CP-violation phases 
\cite{BHP80,Doi81}.
If we identify 
the two independent neutrino mass squared 
differences in this case,
$\Delta m^2_{21}$ and $\Delta m^2_{31}$,  
with the neutrino mass squared differences
which induce the solar and atmospheric 
neutrino oscillations, $\deltasol = \Delta m^2_{21} > 0$,
$\dma = \Delta m^2_{31}$,
one has: $\theta_{12} = \theta_{\odot}$, 
$\theta_{23} = \theta_{\rm atm}$, 
and $\theta_{13} = \theta$. 
The angle $\theta$ is limited by 
the data from the CHOOZ and Palo Verde
experiments~\cite{CHOOZ,PaloV}.
The oscillations
between flavour neutrinos
are insensitive to the Majorana CP-violating phases 
$\alpha_{21}$, $\alpha_{31}$ \cite{BHP80,Lang86}.
Information about these phases can be obtained,
in principle, in the \betabeta-decay experiments 
\cite{BGKP96,BPP1,FViss00,WR00,PPW,PPR1} 
(see also \cite{bb0nuCP1,BargerSNO2bb,deGBorisRabi}).
Majorana CP-violating phases, 
and in particular, the phases $\alpha_{21}$ and/or 
$\alpha_{31}$, might be at the origin of 
the baryon asymmetry of the Universe \cite{LeptoG}. 

  A 3-$\nu$ oscillation analysis of the CHOOZ data
(in this case  
$\Delta m^2 = \dma$)
\cite{BNPChooz} 
led to the conclusion that 
for $\deltasol  \ltap 10^{-4}~{\rm eV^2}$,
the limits on $\sin^2\theta$ 
practically coincide with
those derived in the 2-$\nu$ 
oscillation analysis in \cite{CHOOZ}.
A combined 3-$\nu$ oscillation
analysis of the 
solar neutrino, CHOOZ and the 
KamLAND data, performed
under the assumption 
$\deltasol \ll |\dma|$
(see, e.g., \cite{BGG99,ADE80}), showed 
that \cite{fogli} 
\begin{equation}
\sin^{2} \theta < 0.05,~~~~ 99.73\%~{\rm C.L.}
\label{chooz1}
\end{equation}
%

\noindent 
It was found \cite{fogli} that
the best-fit value of $\sin^2\theta$
lies in the interval 
$\sin^2\theta \cong (0.00 - 0.01)$.

   Somewhat better limits on $\sin^2 \theta$ than 
the existing one can be obtained in the 
MINOS experiment \cite{MINOS}. 
Various options are being currently discussed
(experiments with off-axis neutrino beams, more precise
reactor antineutrino and long baseline experiments, etc.,
see, e.g., \cite{MSpironu02,reactorth13}) of how to improve
by at least a factor of 5 or more, i.e., 
to values of $\sim 0.01$ or smaller, 
the sensitivity to $\sin^2\theta$. 

   Let us note  that the 
atmospheric neutrino and K2K data
do not allow one to determine the sign of 
$\dma$. This implies that if we identify
$\dma$ with $\Delta m^2_{31}$
in the case of 3-neutrino mixing, 
one can have $\Delta m^2_{31} > 0$
or $\Delta m^2_{31} < 0$. The two 
possibilities correspond to two different
types of neutrino mass spectrum:
with normal hierarchy, $m_1 < m_2 < m_3$, and 
with inverted hierarchy,
$m_3 < m_1 < m_2$. 
We will use the terms
{\it normal hierarchical (NH)} and
{\it inverted hierarchical (IH)}
for the two types of spectra in the case of 
strong inequalities between the neutrino masses,
if $m_1 \ll m_2 \ll m_3$ and
\footnote{This definition of the IH
spectrum corresponds to a convention 
we will call A (see, e.g., \cite{HLMACPP03}),
in which the neutrino masses
are not ordered in magnitude
according to their index number.
We can also always number the
neutrinos with definite mass 
in such a way that \cite{BGKP96,BNPChooz}
$m_1 < m_2 < m_3$. In this convention
called B \cite{HLMACPP03}
the IH spectrum corresponds to
$m_1 << m_2 \cong m_3$. We will use 
convention B in our further analysis.}
$m_3 \ll m_1 < m_2$, respectively.
The spectrum can also be of
{\it quasi-degenerate (QD)} type:
$m_1 \cong m_2 \cong m_3$,
$m^2_{1,2,3} >> |\dma|$. 

   The sign of \dma can be determined in
very long base-line neutrino oscillation 
experiments at neutrino factories
(see, e.g., \cite{AMMS}), and, e.g, using
combined data from long base-line
oscillation experiments at the JHF facility and
with off-axis neutrino beams \cite{HLM}.
Under certain rather special
conditions it might be determined also
in experiments with reactor
$\bar{\nu}_e$ \cite{SPMPiai01,HLMACPP03}.

   As is well-known, neutrino 
oscillation experiments allow one to determine
differences of squares of neutrino masses,
but not the absolute values of
the masses. Information on the absolute 
values of neutrino masses
of interest can be derived in the \hbeta 
experiments studying the 
electron spectrum \cite{Fermi34,MoscowH3,MainzKATRIN}
and from cosmological and astrophysical 
data (see, e.g., ref.~\cite{MAPPLANCK,WMAPnu,Hu99}).
The currently existing most stringent upper 
bounds on the electron (anti-)neutrino mass  
$m_{\bar{\nu}_e}$ were obtained in the
Troitzk~\cite{MoscowH3} and Mainz~\cite{MainzKATRIN} 
\hbeta experiments and read:
\begin{eqnarray}
m_{\bar{\nu}_e}  <  2.2 \eV ~~~(95\%~C.L.).
\label{H3beta}
\end{eqnarray}
%
\noindent 
We have $m_{\bar{\nu}_e} \cong m_{1,2,3}$
in the case of QD neutrino mass spectrum.
The KATRIN \hbeta experiment \cite{MainzKATRIN}
is planned to reach a sensitivity  to  
$m_{\bar{\nu}_e} \sim (0.20 - 0.35)$ eV. 
The data of the WMAP experiment  
on the cosmic microwave background 
radiation was used to 
obtain an upper limit on the sum of the neutrino 
masses \cite{WMAPnu}: 
\begin{eqnarray}
\sum_{j} m_{j} < 0.70~ \eV ~~~(95\%~C.L.).
\label{WMAP}
\end{eqnarray}
%
\noindent A conservative 
estimate of all the uncertainties 
related to the derivation of this 
result (see, e.g., \cite{Hanne03})
lead to a less stringent
upper limit at least by a factor 
of $\sim 1.5$ and possibly 
by a factor of $\sim 3$.
The WMAP and future PLANCK 
experiments can be sensitive to 
\cite{MAPPLANCK} $\sum_{j} m_{j}  \cong  0.4$ eV.
\noindent Data on weak lensing of 
galaxies by large scale structure,
combined with data from the WMAP and PLANCK 
experiments may allow one to determine 
$(m_1 + m_2 + m_3)$ with an uncertainty
of \cite{Hu99} $\delta \sim  0.04$ eV.

  After the spectacular experimental 
progress made in the last two years or so 
in the studies of neutrino oscillations, 
further understanding
of the structure of the neutrino masses 
and mixing, of their origins  
and of the status of the CP-symmetry in 
the lepton sector requires a large and
challenging program of 
research to be pursued 
in neutrino physics. 
The main goals of such a 
research program should include:\\
-- High precision determination of 
neutrino mixing parameters
which control the 
solar and the dominant atmospheric  
neutrino oscillations, 
$\deltasol$, $\theta_{\odot}$, and 
$\dma$, $\theta_{atm}$.\\
-- Measurement of, or improving by 
at least a factor of (5 - 10)
the existing upper limit on, the value of 
the only small mixing angle
$\theta~(=\theta_{13})$ 
in the PMNS matrix $U$.\\
-- Determination of the 
type of the neutrino mass spectrum 
(normal hierarchical,
or inverted hierarchical, 
or quasi-degenerate).\\
-- Determining or obtaining significant constraints
on the absolute scale of neutrino masses, or on the lightest 
neutrino mass.\\ 
-- Determining the nature
of massive neutrinos  which can be Dirac 
or Majorana particles.\\
-- Establish whether the CP-symmetry is violated in the lepton 
sector a) due to the Dirac phase $\delta$, and/or
b) due to the Majorana phases $\alpha_{21}$ and $\alpha_{31}$
if the massive neutrinos are Majorana particles.\\
-- Searching with increased sensitivity
for possible manifestations, other than 
flavour neutrino oscillations,
of the non-conservation
of the individual lepton charges $L_l$, $l=e,\mu,\tau$,
such as $\mu \rightarrow e + \gamma$,
$\tau \rightarrow \mu + \gamma$, etc. decays.\\
-- Understanding at fundamental level 
the mechanism giving rise to the 
neutrino masses and mixing and to the
$L_l-$non-conservation,
i.e., finding the Theory of neutrino
mixing. Progress in 
the theory of neutrino mixing
might also lead, in particular, 
to a better understanding of the 
possible relation between 
CP-violation in the lepton sector 
at low energies and the generation 
of the baryon asymmetry of the Universe.

  Obviously, the successful realization 
of the experimental part of this program of 
research would be a formidable task
and would require most probably (10 - 15) years.   

    In the present article
we will review the potential contribution
the studies of neutrinoless double 
beta ($\betabeta-$) decay of certain even-even
nuclei, $(A,Z) \rightarrow (A,Z + 2) + e^- + e^-$, 
can make to the program of research
outlined above. The \betabeta-decay is allowed 
if the neutrinos with definite mass 
are Majorana particles (for reviews see, 
 e.g., \cite{BiPet87,ElliotVogel02}).
Let us recall that the nature - Dirac or Majorana,
of the massive neutrinos 
$\nu_j$, 
is related to the fundamental symmetries of the 
particle interactions. The neutrinos $\nu_j$
will be Dirac fermions if the 
particle interactions conserve some 
lepton charge, which could be, e.g., the total
lepton charge $L$. The neutrinos with 
definite mass can be 
Majorana particles
if there does not exist 
any conserved lepton charge.
As is well-known, the massive neutrinos are 
predicted to be of Majorana nature
by the see-saw mechanism 
of neutrino mass generation~\cite{seesaw},
which also provides a  
very attractive explanation of the
smallness of the neutrino masses 
and - through the leptogenesis theory 
\cite{LeptoG},
of the observed baryon asymmetry 
of the Universe.

  If the massive neutrinos $\nu_j$ 
are Majorana fermions, 
processes in which the total lepton charge
$L$ is not conserved and changes by two units, 
such as $K^+ \rightarrow \pi^- + \mu^+ + \mu^+$,
$\mu^+ + (A,Z) \rightarrow (A,Z+2) + \mu^-$, etc.,
should exist. The process most sensitive to the 
possible Majorana nature of the  
massive neutrinos $\nu_j$ is  
the $\betabeta-$decay
(see, e.g., \cite{BiPet87}).
If the \betabeta-decay is generated
{\it only by the (V-A) charged current 
weak interaction via the exchange of the three
Majorana neutrinos  $\nu_j$ and the latter
have masses not exceeding few MeV},
which will be assumed to hold throughout
this article, the dependence of the 
\betabeta-decay amplitude $A\betabeta$ 
on the neutrino mass and mixing parameters
factorizes in the 
effective Majorana mass \mefff{}
(see, e.g., \cite{BiPet87,ElliotVogel02}):
\begin{equation}
A\betabeta \sim \mefff~M~,
\label{Abb}
\end{equation}
%
\noindent where $M$ is the corresponding 
nuclear matrix element (NME) and
\begin{equation}
\meff  = \left| m_1 |U_{\mathrm{e} 1}|^2 
+ m_2 |U_{\mathrm{e} 2}|^2~e^{i\alpha_{21}}
 + m_3 |U_{\mathrm{e} 3}|^2~e^{i\alpha_{31}} \right|~,
\label{effmass2}
\end{equation}
\noindent
$\alpha_{21}$ and $\alpha_{31}$ 
being the two Majorana CP-violating phases
of the PMNS matrix 
\footnote{We assume that $m_j > 0$ and that
the fields of the 
Majorana neutrinos $\nu_j$ 
satisfy the Majorana condition:
$C(\bar{\nu}_{j})^{T} = \nu_{j},~j=1,2,3$,
where $C$ is the charge conjugation matrix.}
\cite{BHP80,Doi81}. Let us note
that if CP-invariance holds, 
one has \cite{LW81}
$\alpha_{21} = k\pi$, $\alpha_{31} = 
k'\pi$, where $k,k'=0,1,2,...$. In this case 
\begin{equation}
\eta_{21} \equiv e^{i\alpha_{21}} = \pm 1,~~~
\eta_{31} \equiv e^{i\alpha_{31}} = \pm 1 
\label{eta2131}
\end{equation}
\noindent represent the relative 
CP-parities of the Majorana neutrinos 
$\nu_1$ and $\nu_2$, and 
$\nu_1$ and $\nu_3$, respectively.
It follows from eq. (\ref{effmass2}) that
the measurement of $\meff$ will provide
information, in particular, 
on the neutrino masses.
As eq. (\ref{Abb}) indicates,
the observation of the 
\betabeta-decay of a given nucleus and
the measurement of the corresponding
half life-time, would allow one to determine 
$\meff$ only if the value of the relevant
NME $M$ is known 
with a relatively small uncertainty.

   The experimental searches for $\betabeta-$decay
have a long history (see, e.g., \cite{ElliotVogel02}).
Rather stringent upper bounds on 
\meff{} have been obtained in the 
$^{76}$Ge experiments 
by the Heidelberg-Moscow 
collaboration \cite{76Ge00}: 
\begin{equation}
\meff < 0.35 \ \mathrm{eV},~~~90\%~{\rm C.L.}
\label{76Ge0}
\end{equation}
%
\noindent Taking into account a factor of 3 
uncertainty associated with 
the calculation of the relevant
nuclear matrix element 
\cite{ElliotVogel02}, we get
\footnote{Evidences for \betabeta-decay
taking place with a rate corresponding to
$0.11 \ {\rm eV} \leq  \meff  \leq  0.56$ eV
(95\% C.L.) are claimed to 
have been obtained in \cite{Klap01}. The
results announced in \cite{Klap01} have been 
criticized in \cite{bb0nu02}.
}
\begin{equation}
\meff < (0.35 - 1.05) \ \mathrm{eV},~~~90\%~{\rm C.L.}
\label{76Ge00}
\end{equation}
%
\noindent The IGEX collaboration  
has obtained \cite{IGEX00}:
\begin{equation}
\meff < (0.33 \div 1.35) \ \mathrm{eV},~~~90\%~{\rm C.L.}
\label{IGEX00}
\end{equation}

   Higher sensitivity to 
\meff{} is planned to be 
reached in several $\betabeta$-decay experiments
of a new generation. 
The NEMO3 experiment \cite{NEMO3} 
with $^{100}$M and $^{82}$Se,
which began to take data in July of 2002, 
and the cryogenics detector CUORICINO 
\cite{CUORE}, which uses 
$^{130}$Te and
is already operative, 
are expected to reach 
a sensitivity to values of 
$\meff  \sim 0.2~$eV.
The first preliminary results from 
these two experiments were
announced recently \cite{NEMO3,CUORICINO}
and respectively read (90\% C.L.):
$\meff  < (1.2 - 2.7)$ eV and
$\meff < (0.7 - 1.7)$ eV.
Up to an order of magnitude better sensitivity, 
i.e., to $\meff  \cong 2.7\times 10^{-2}$ eV,
$1.5\times 10^{-2}~$eV, $5.0\times 10^{-2}~$eV,
$2.5\times 10^{-2}$ eV and $3.6\times 10^{-2}$ eV
is planned to be achieved 
in the CUORE, 
GENIUS,
EXO,
MAJORANA
and MOON
experiments \cite{CUORE} 
\footnote{The quoted sensitivities 
correspond to values of the relevant NME
from ref.\ \cite{SMutoKK90}.}
with $^{130}$Te, $^{76}$Ge, $^{136}$Xe,
$^{76}$Ge and $^{100}$Mo, respectively.
Additional high sensitivity experiments with
$^{136}$Xe - XMASS \cite{XMASS}, and with
$^{48}$Ca - CANDLES \cite{CANDLES}, are also being 
considered.

   As we will discuss in what follows,
the studies of $\betabeta-$decay and
a measurement of a nonzero value of 
$\meff \gtap {\rm few}~10^{-2}$ eV:\\
-- Can establish the Majorana 
nature of massive neutrinos. The \betabeta-decay 
experiments are presently 
the only feasible experiments capable of
doing that (see \cite{BiPet87}).\\
-- Can give information on the 
type of neutrino mass spectrum 
\cite{BGGKP99,BG99,BPP1,PPW,WR00,bb0nuMassSpec1,PPSNO2bb,PPRSNO2bb}.
More specifically, a measured value of 
$\meff \sim {\rm few}\times 10^{-2}$ eV
can provide, in particular,
unique constraints on,
or even can allow one to determine, 
the type of the neutrino mass spectrum 
if $\nu_{1,2,3}$ are Majorana particles
\cite{PPSNO2bb}.\\
-- Can provide also unique information on the 
absolute scale of neutrino masses, or on the 
lightest neutrino mass (see, e.g., 
\cite{BGGKP99,BG99,PPW,bb0nuMassSpec1}). \\ 
-- With additional information
from other sources (\hbeta  experiments
or cosmological and astrophysical 
data and considerations)
on the absolute neutrino mass scale,
the $\betabeta-$decay experiments
can provide unique information
on the Majorana CP-violation phases
$\alpha_{21}$ and $\alpha_{31}$ 
\cite{BGKP96,BPP1,WR00,PPW,PPR1}.

\vspace{-0.3cm}
\section{Predictions for the Effective Majorana Mass 
}
\vspace{-0.2cm}

The predicted value 
of \meff{} depends in the 
case of $3-\nu$ mixing on~\cite{SPAS94}
(see also \cite{BGGKP99,BPP1}):
i) \dma$\!\!$,
ii) $\theta_{\odot}$ and $\Delta m^2_{\odot}$, 
iii) the lightest neutrino mass, 
and on iv) the mixing angle $\theta$.
Using the convention (B) 
in which always $m_1 < m_2 < m_3$, 
one has $\dma \equiv \Delta m^2_{31}$,
and $m_3 = \sqrt{m_1^2 + \dma}$, while
either $\deltasol \equiv \Delta m^2_{21}$ 
(normal mass hierarchy)
or $\deltasol \equiv \Delta m^2_{32}$
(inverted mass hierarchy).
In the first case we have 
$m_2 = \sqrt{m_1^2 + \deltasol}$,
$|U_{\mathrm{e} 1}|^2 = \cos^2\theta_{\odot} (1 - |U_{\mathrm{e} 3}|^2)$, 
$|U_{\mathrm{e} 2}|^2 = \sin^2\theta_{\odot} (1 - |U_{\mathrm{e} 3}|^2)$,
and  $|U_{\mathrm{e} 3}|^2 \equiv \sin^2\theta$,
while in the second 
$m_2 = \sqrt{m_1^2 + \dma - \deltasol}$,
$|U_{\mathrm{e} 2}|^2 = \cos^2\theta_{\odot} (1 - |U_{\mathrm{e} 1}|^2)$, 
$|U_{\mathrm{e} 3}|^2 = \sin^2\theta_{\odot} (1 - |U_{\mathrm{e} 1}|^2)$,
and  $|U_{\mathrm{e} 1}|^2 \equiv \sin^2\theta$.
The two possibilities for \deltasol{}
correspond also to the 
two different 
{\it hierarchical} types
of neutrino mass spectrum ---  
the {\it normal hierarchical~(NH)},
$m_1 << m_2 << m_3$, and
the {\it inverted hierarchical (IH)},
$m_1 << m_2 \cong m_3$, respectively.
Let us recall that in the case of
{\it quasi-degenerate (QD)} 
neutrino mass spectrum we have
$m_1 \cong m_2 \cong m_3$,
$m^2_{1,2,3} >> \dma$.
For the  allowed ranges of values of 
\deltasol{} and $\dma$ \cite{SKatm00}, 
the NH (IH) spectrum
corresponds to $m_1 \ltap 10^{-3}~(2\times 10^{-2})$ eV,
while one has a QD spectrum
if $m_{1,2,3}\cong m_{\bar{\nu}_e} > 0.20$ eV.
For $m_1$ lying in the interval
between $\sim 10^{-3}~(2\times 10^{-2})$ eV
and 0.20 eV, the neutrino mass spectrum
is with partial normal (inverted)
hierarchy (see, e.g., \cite{BPP1}).

     Given \dms$\!\!$, \dma$\!\!$, 
$\theta_{\odot}$ and
$\sin^2\theta$, the value of \meff{} 
depends strongly on the type of the
neutrino mass spectrum as well as 
on the values of the two
Majorana CP-violation phases
of the PMNS matrix,
$\alpha_{21}$ and $\alpha_{31}$ 
(see eq.\ (\ref{effmass2})).
Let us note that in the case 
of QD spectrum,
$m_1 \cong m_2 \cong m_3$, 
$m_{1,2,3}^2 \gg \dma\!\!, \deltasol$,
~\meff{} is essentially independent on
\dma and \deltasol, and 
the two possibilities, 
$\deltasol \equiv \deltadue$
and $\deltasol \equiv \deltatre \!\!$, 
lead {\it effectively} 
to the same predictions for  
\meff{} 
\footnote{This statement is valid, 
within the convention 
$m_1 < m_2 < m_3$ we are using,
as long as there are no independent
constraints on the  CP-violating phases
$\alpha_{21}$ and $\alpha_{31}$ 
which enter into the expression for \meff.
In the case of spectrum with normal hierarchy, 
\meff{}  depends primarily 
on $\alpha_{21}$ ($|U_{\mathrm{e} 3}|^2 \ll 1$), 
while if the spectrum is with 
inverted hierarchy,
\meff{}  will depend essentially
on $\alpha_{31} - \alpha_{21}$ 
($|U_{\mathrm{e} 1}|^2 \ll 1$).}.

\vspace{-0.3cm}
\subsection{Normal Hierarchical Neutrino Mass Spectrum}
\vspace{-0.2cm}

In the case of NH neutrino mass spectrum one has
$m_2  \cong  \sqrt{\deltasol}$,
$m_3 \cong  \sqrt{\deltaatm}$, 
$|U_{\mathrm{e} 3}|^2 \equiv \sin^2\theta$,
and correspondingly
\begin{equation}
\meff = \left|(m_1 \cos^2\theta_\odot  + 
e^{i\alpha_{21}} \sqrt{\deltasol}
\sin^2 \theta_\odot)\cos^2\theta \right.
+ \left. \sqrt{\dma} \sin^2\theta~e^{i\alpha_{31}} \right| 
\label{meffNH1}
\end{equation}
\begin{equation}
\simeq \left| \sqrt{\deltasol} 
\sin^2 \theta_\odot \cos^2\theta + \sqrt{\dma} 
\sin^2\theta e^{i(\alpha_{31} - \alpha_{21})} \right| \nonumber
\label{meffNH2}
\end{equation}

\noindent where 
we have neglected the 
term  $\sim m_1$ in eq. (\ref{meffNH2}). 
Although in this case
one of the three massive Majorana 
neutrinos effectively
``decouples'' and does not give a contribution to 
\meff, the value of \meff
still depends on the Majorana CP-violation 
phase $\alpha_{32} = \alpha_{31} - \alpha_{21}$.
This reflects the fact that in contrast
to the case of massive Dirac 
neutrinos (or quarks),
CP-violation can take place
in the mixing of only two massive 
Majorana neutrinos \cite{BHP80}.   

  Since, as it follows from 
eqs. (\ref{sol90}) and (\ref{atmo90}), we have
$\sqrt{\deltasol} \ltap 1.3\times 10^{-2}$ eV,
$\sin^2\theta_{\odot} \ltap 0.42$,
$\sqrt{\dma} \ltap 5.5\times 10^{-2}$ eV,
and the largest neutrino mass
enters into the expression for $\meff$ with
the factor $\sin^2\theta < 0.05$, 
the predicted value of $\meff$ is below
$10^{-2}$ eV: 
for $\sin^2\theta = 0.05~(0.01)$ one finds
$\meff \ltap 0.0086~(0.0066)$ eV.
Using the best fit values of 
the indicated neutrino oscillation 
parameters we get even 
smaller values for $\meff$,
$\meff \ltap 0.0059~(0.0039)$ eV
(see Tables 1 and 2). Actually, 
it follows from eq. (\ref{meffNH1}) 
and the allowed ranges of values of 
$\deltasol$, $\dma$,
$\sin^2\theta_{\odot}$,
$\sin^2\theta$ as well as 
of the lightest neutrino mass $m_1$ and
the CP-violation phases 
$\alpha_{21}$ and  $\alpha_{31}$,
that in the case of NH spectrum 
there can be a complete 
cancellation between the contributions of the
three terms in eq. (\ref{meffNH1}) and
one can have \cite{PPW} $\meff = 0$.
 
\vspace{-0.3cm}
\subsection{Inverted Hierarchical 
Spectrum}
\vspace{-0.2cm}
 
 One has for the IH neutrino mass spectrum 
(see, e.g. \cite{BPP1}) 
$m_2 \cong m_3 \cong \sqrt{\dma}$, 
$|U_{\mathrm{e} 1}|^2 \equiv \sin^2\theta$.
 Neglecting 
$m_1 \sin^2\theta$ in eq. (\ref{effmass2}),
we find \cite{BGKP96,BGGKP99,BPP1}:
\begin{equation}
\meff \cong \sqrt{\dma} \cos^2\theta \sqrt{1 - \sin^22\theta_{\odot} 
\sin^2\left (\frac{\alpha_{32}}{2} \right)}.
\label{meffIH1}
\end{equation}
%
\noindent Even though one of the
three massive Majorana neutrinos 
``decouples'', the value of $\meff$
depends on the Majorana CP-violating phase
$\alpha_{32} \equiv (\alpha_{31} - \alpha_{21})$. 
Obviously, 
\begin{equation}
\sqrt{\dma}\cos^2\theta|~\cos 2 \theta_\odot|~
\leq~ \meff \leq \sqrt{\dma}\cos^2\theta.
\label{meffIH2}
\end{equation}
%
\noindent  The upper and the lower limits
correspond respectively to the 
CP-conserving cases $\alpha_{32} = 0$, or
$\alpha_{21} = \alpha_{31} = 0, \pm \pi$, 
and $\alpha_{32} = \pm \pi$,
or $\alpha_{21} = \alpha_{31} + \pi = 0, \pm  \pi$.
Most remarkably,
since according to the 
solar neutrino and KamLAND data
$\cos 2 \theta_\odot \sim (0.35 - 0.40)$,
we get a significant lower limit on 
$\meff$, typically exceeding $10^{-2}$ eV,
in this case \cite{PPSNO2bb,PPW}
(Tables 1 and 2). Using, e.g., the best 
fit values of $\dma$ and 
$\tan^2\theta_{\odot}$ one finds:
$\meff \gtap 0.018$ eV.
The maximal value of $\meff$ is determined 
by $\dma$ and can reach, as it follows from
eq. (\ref{atmo90}), $\dma \sim 6\times 10^{-2}$ eV.
The indicated values of \meff are within 
the range of sensitivity of the next generation of
\betabeta-decay experiments.

  The expression for \meff, eq. (\ref{meffIH1}),
permits to relate the value of
$\sin^2 (\alpha_{31} - \alpha_{21})/2$ 
to the experimentally 
measured quantities \cite{BGKP96,BPP1}
\meff, \deltaatm and $\sin^22\theta_{\odot}$: 
\begin{equation}
\sin^2  \frac{\alpha_{31} - \alpha_{21}}{2}  \cong
\left( 1 - \frac{\meff^2}{\dma \cos^4\theta} \right) 
\frac{1}{\sin^2 2 \theta_\odot}.
\end{equation}
%
\noindent A more precise 
determination of $\dma$
and  $\theta_\odot$ 
and a sufficiently accurate measurement of \meff 
could allow one to get information 
about the value of 
$(\alpha_{31} - \alpha_{21})$,
provided the neutrino mass spectrum 
is of the IH type.

\vspace{-0.3cm}
\subsection{Three Quasi-Degenerate Neutrinos}
\vspace{-0.3cm}

 In this case it is convenient to introduce
$m_0 \equiv m_1 \cong  m_2 \cong m_3$,
$m^2_0 \gg \dma$, $m_0 \gtap 0.20$ eV. 
The mass $m_0$ effectively coincides with 
the electron (anti-)neutrino mass $m_{\bar{\nu}_e}$ 
measured in the $^{3}$H $\beta$-decay experiments:
$m_0 = m_{\bar{\nu}_e}$.   
Thus, $m_0 < 2.2 \eV$, or if we use a
conservative cosmological upper limit \cite{Hanne03}
$m_0 < 0.7$ eV.  The QD neutrino mass 
spectrum is realized for values of  
$m_0$, which can be
measured in the $^3$H $\beta-$decay experiment KATRIN.

 The effective Majorana mass \meff is given by
\begin{equation}
\meff \cong m_0 \left| (\cos^2 \theta_\odot
 + ~\sin^2 \theta_\odot 
 e^{i \alpha_{21}})~\cos^2\theta + 
 e^{i \alpha_{31}} \sin^2\theta \right| 
\label{meffQD0}
\end{equation}
\begin{equation}
\cong m_0 \left| \cos^2 \theta_\odot + 
~\sin^2 \theta_\odot e^{i \alpha_{21}} \right|
= m_0~ \sqrt{1 - \sin^22\theta_{\odot} 
\sin^2\left (\frac{\alpha_{21}}{2} \right)}.  
\label{meffQD1}
\end{equation}
%
\noindent Similarly to the case of IH spectrum, one has:
\begin{equation}
m_0~\left| \cos2\theta_\odot \right|~ \ltap \meff \ltap m_0. 
\label{meffQD2}
\end{equation}
%
\noindent For $\cos 2 \theta_\odot \sim (0.35 - 0.40)$
favored by the solar neutrino and the KamLAND data
one finds a non-trivial lower limit of
on \meff, $\meff \gtap (0.06 - 0.07)$ eV.
Using the conservative cosmological upper bound on
the sum of neutrino masses 
we get $\meff \ltap 0.70$ eV.
Also in this case one can obtain, 
in principle, a direct information
on one CP-violation phase
from the measurement of $\meff$,
$m_0$ and $\sin^2 2 \theta_\odot$:
\begin{equation}
\sin^2  \frac{\alpha_{21}}{2}  \cong 
\left( 1 - \frac{\meff^2}{m^2_0} \right) 
\frac{1}{\sin^2 2 \theta_\odot}.
\end{equation}

\vspace{0.4cm}
  The specific features of the predictions for \meff
in the cases of the three types of neutrino mass 
spectrum discussed above are evident 
in Figs. 1 and 2, where the
dependence of \meff
on $m_1$ for the two possible sub-regions of the 
LMA solution region - LMA-I and LMA-II,
is shown. If $\deltasol = \Delta m^2_{21}$,
for instance, which corresponds to
a spectrum with normal hierarchy, 
\meff can lie anywhere
between 0 and the presently existing upper limits, 
given by eqs. (\ref{76Ge00}) and (\ref{IGEX00}).
This conclusion does not change even 
under the most favorable
conditions for the determination of \meff,
namely, even when \deltaatm, \deltasol,
$\theta_{\odot}$ and $\theta$ are known
with negligible uncertainty, as Fig. 1, upper left panel, and 
Fig. 2, upper panel, indicate.

\vspace{-0.4cm}
\section{Constraining 
the Lightest Neutrino Mass}
\vspace{-0.3cm}

 If the \betabeta-decay of a given nucleus 
will be observed, it would be possible to 
determine the value of \meff from the
measurement of the associated life-time of the decay.
This would require the knowledge of the nuclear
matrix element of the process. At present there
exist large uncertainties in the calculation of
the \betabeta-decay nuclear matrix elements
(see, e.g., \cite{ElliotVogel02,bilgri}).
This is reflected, in particular, in the
factor of $\sim 3$ uncertainty
in the upper limit on \meff , which is
extracted from the experimental
lower limits on the \betabeta-decay 
half life-time of $^{76}$Ge. The observation of
a \betabeta-decay of one nucleus is likely to lead
to the searches and eventually to observation
of the decay of other nuclei.
One can expect that such a progress, in particular,
will help to solve the problem 
of the sufficiently precise calculation
of the nuclear matrix elements 
for the \betabeta-decay.

    In this Section we consider briefly 
the information that future 
\betabeta-decay and/or $^3$H $\beta-$decay 
experiments can provide on the 
lightest neutrino mass $m_1$,
without taking into account
the possible effects 
of the currently existing
uncertainties in the evaluation
of the \betabeta-decay nuclear matrix elements.

  An experimental upper 
limit on \meff, $\meff < \meff_{exp}$,
will determine a maximal value of $m_1$,
$m_1 < (m_1)_{max}$ in the case of 
normal mass hierarchy, 
$\deltasol \equiv \deltadue$ (Figs. 1, 2). 
For the QD 
spectrum, for instance,
we have $m_1 \gg \deltasol,\dma$,
and up to corrections 
$\sim \deltasol \sin^2\theta_\odot /(2m_1^2)$
and $\sim \dma \sin^2\theta/(2m_1^2)$ one finds 
\cite{PPW,Poles2}:
\begin{equation}
 (m_1)_{max} \cong \frac{\meff_{exp}}
{\left|\cos 2\theta_\odot\cos^2\theta
- \sin^2\theta \right|}~.
\label{maxm1nhLMAQD}
\end{equation}
\noindent We get similar results in the case of 
inverted mass hierarchy, $\deltasol \equiv \deltatre$,
provided the experimental upper limit 
$\meff_{exp}$ is larger than the 
minimal value of \meff,  
$\meff_{min}^{ph}$ (Figs. 1, 2), 
predicted by taking into account 
all uncertainties in the values of 
the relevant input parameters 
(\dma, \deltasol, $\theta_{\odot}$, etc.).
If $\meff_{exp} < \meff_{min}^{ph}$, 
then either 
i) the neutrino mass spectrum is not
of the inverted hierarchy type, or ii) 
there exist contributions to
the \betabeta-decay rate other than 
due to the light Majorana neutrino exchange
(see, e.g., \cite{bb0nunmi}) that 
partially cancel the
contribution from the 
Majorana neutrino exchange. 
The indicated result might also 
suggest that the massive neutrinos 
are Dirac particles.

 A measurement  of $\meff = (\meff)_{exp}
\gtap 0.02$ eV if $\deltasol \equiv \deltadue$,
and of $\meff = (\meff)_{exp}
\gtap \sqrt{\dma}\cos^2\theta$
in the case of $\deltasol \equiv \deltatre$,
would imply that $m_1 \gtap 0.02$ eV and 
$m_1 \gtap 0.04$ eV, respectively,
and thus a neutrino mass spectrum 
with partial hierarchy
or of the QD type \cite{BPP1} (Figs. 1, 2). 
The lightest neutrino mass
will be constrained to lie in a
rather narrow interval,
$(m_1)_{min} \leq m_1 \leq (m_1)_{max}$
\footnote{Analytic expressions for
$(m_1)_{min}$ and $(m_1)_{max}$
are given in \cite{PPW}.}. 
The limiting values of $m_1$ 
correspond to the case of CP-conservation.
For $\deltasol \ll m_1^2$, 
(i.e., for $\deltasol \ltap 10^{-4}~{\rm eV^2}$),
as can be shown \cite{PPW}, 
we have $(m_1)_{min} \cong (\meff)_{exp}$
for $\deltasol \equiv \deltadue$,
and $\sqrt{((m_1)_{min})^2 + \dma}~
\cong (\meff)_{exp}$
for $\deltasol \equiv \deltatre$.

 A measured value of \meff satisfying 
$(\meff)_{exp} < (\meff)_{max}$, where, 
e.g., in the case of a QD spectrum 
$(\meff)_{max} \cong m_1 \cong m_{\bar{\nu}_e}$,
would imply that 
at least one of the two 
CP-violating phases
is different from zero :
$\alpha_{21}\neq 0$ or $\alpha_{31} \neq 0$
\footnote{Let us note that,
in general, the knowledge of the value
of \meff alone will not allow to 
distinguish the case
of CP-conservation 
from that of CP-violation.}.

  If the measured value of \meff lies 
between the minimal and maximal values of
\meff, predicted in the case of 
inverted {\it hierarchical} spectrum, 
\begin{equation}
\meff_{\pm} = 
\left |\sqrt{\dma - \deltasol}
\cos^2 \theta_\odot \pm \sqrt{\dma}
\sin^2 \theta_\odot \right | \cos^2\theta, 
\label{0m1ih}
\end{equation}
%
\noindent $m_1$ again would be limited from above,
but we would have $(m_1)_{min} = 0$
(Figs. 1, 2).

  A measured value of $m_{\bar{\nu}_e}$,
$(m_{\bar{\nu}_e})_{exp} \gtap 0.20$ eV, 
satisfying $(m_{\bar{\nu}_e})_{exp} > (m_1)_{max}$,
where $(m_1)_{max}$ is determined 
from the upper limit on \meff
in the case the \betabeta-decay is not 
observed, might imply 
that the massive neutrinos are Dirac particles.
If \betabeta-decay has been observed and \meff
measured, the inequality
$(m_{\bar{\nu}_e})_{exp} > (m_1)_{max}$, with
$(m_1)_{max}$ determined from 
the measured value of \meff,
would lead to the conclusion that
there exist contribution(s) to
the \betabeta-decay rate other than 
due to the light Majorana neutrino exchange
(see, e.g., \cite{bb0nunmi} 
and the references quoted therein) that 
partially cancels the
contribution from the 
Majorana neutrino exchange.

\vspace{-0.3cm}
\section{Determining the Type of Neutrino Mass Spectrum} 
\vspace{-0.3cm}

   The possibility to distinguish
between the three different types 
of neutrino mass spectrum - NH, IH and QD,
depends on the allowed ranges of 
values of \meff{} for the 
three spectra. More specifically, 
it is determined by 
the maximal values of \meff{} 
in the cases of NH and IH spectra,
$\meff_{\rm max}^{\rm NH}$ and 
$\meff_{\rm max}^{\rm IH}$,
and by the minimal values of \meff{} 
for the IH and QD spectra,
$ \meff_{\rm min}^{\rm IH}$ and $\meff_{\rm min}^{\rm QD}$.
These can be derived from eqs. (\ref{meffNH1}),
(\ref{meffIH2}) and (\ref{meffQD0}) and correspond
to CP-conserving values of the Majorana phases \cite{PPRSNO2bb}
$\alpha_{21}$ and  $\alpha_{31}$.
The minimal value $\meff_{\rm min}^{\rm QD}$
scales to a good approximation with $m_0$
and thus is reached for $m_0 = 0.2$ eV.

 In Tables 1 and 2 (taken from \cite{PPRSNO2bb}) 
we show the values of 
i) $\meff_{\rm max}^{\rm NH}$,
ii) $ \meff_{\rm min}^{\rm IH}$,
and iii) $\meff_{\rm min}^{\rm QD}$
($m_0 = 0.2$ eV), calculated 
for the best-fit and
the 90\% C.L.\ 
allowed ranges of
values of $\tan^2\theta_{\odot}$ and \deltasol~
in the LMA solution region.
In Table 3 (from \cite{PPRSNO2bb})
we give the same quantities,
$\meff_{\rm max}^{\rm NH}$,
$\meff_{\rm min}^{\rm IH}$
and $\meff_{\rm min}^{\rm QD}$,
calculated using the best-fit values
of the neutrino oscillation parameters,
including 1 s.d. (3 s.d.) 
``prospective'' uncertainties 
\footnote{For further details
concerning the calculation
of the uncertainty in \meff
in this case see \cite{PPR1,PPRSNO2bb}.} 
of 5\% (15\%) on $\tan^2\theta_{\odot}$
and $\deltasol$, and of 10\% (30\%) on
\dma$\!\!$. As is evident from 
Tables 1 - 3, the possibility 
of determining the type of the
neutrino mass spectrum
if \meff{} is found to be nonzero in the 
\betabeta-decay experiments of the next 
generation, depends crucially  
on the precision with which 
\dma$\!\!$, $\theta_{\odot}$, \deltasol, 
 $\sin^2\theta$ and \meff{} 
will be measured. It depends also crucially
on the values of
$\theta_{\odot}$ and of \meff.
The precision itself of the measurement of 
\meff{} in the next generation of \betabeta-
decay experiments, given the latter sensitivity
limits of $\sim (1.5 - 5.0)\times 10^{-2}~{\rm eV}$,
depends on the value of \meff.   
The precision 
in the measurements 
of $\tan^2\theta_{\odot}$
and $\deltasol$  used in order 
to derive the numbers in Table 3
can be achieved, e.g., 
in the solar neutrino
experiments and/or in the
experiments with reactor $\bar{\nu}_e$ 
\cite{Sandh70,HLMACPP03}.
If \dma lies in the interval 
$\dma \!\!\cong (2.0 - 5.0)\times 10^{-3}~{\rm eV^2}$, 
as is suggested by the current 
data \cite{SKatm00,fogliold}, 
its value will be determined with a 
$\sim 10\%$ error (1 s.d.)  
by the MINOS experiment \cite{MINOS}.

  The high precision measurements of 
\dma$\!\!$, $\tan^2\theta_{\odot}$ and
\deltasol~are expected to take place 
within the next $\sim (6 - 7)$ years. 
We will assume in what follows that 
the problem of measuring
or tightly constraining $\sin^2\theta$ will also be  
resolved  within the indicated period.
Under these conditions, 
the largest uncertainty
in the comparison of the theoretically 
predicted value of \meff{} with that 
determined in the
\betabeta-decay experiments would be associated
with the corresponding \betabeta-decay
nuclear matrix elements. 
We will also assume in what follows that
by the time one or more \betabeta-decay 
experiments of the next generation
will be operative ($2009 - 2010$)
at least the physical range
of variation of the values of the relevant
\betabeta-decay nuclear matrix elements 
will be unambiguously determined.

   Following \cite{PPR1,PPRSNO2bb}, 
we will parametrize the 
uncertainty in \meff{}
due to the  imprecise knowledge of the 
relevant nuclear matrix elements ---
we will use the term ``theoretical uncertainty'' 
for the latter --- through a parameter $\zeta$, 
$\zeta \geq 1$, defined as:
\begin{equation} \label{eq:zeta}
\meff = \zeta \Big( (\meff_{\rm exp})_{\mbox{}_{\rm MIN}} \pm \Delta
\Big)~,
\end{equation}
where $(\meff_{\rm exp})_{\mbox{}_{\rm MIN}}$ 
is the value of \meff{}
obtained from the measured 
\betabeta-decay half life-time
of a given nucleus using 
{\it the largest nuclear matrix element}
and $\Delta$ is  the experimental error. 
An experiment measuring a \betabeta-decay 
half-life time 
will thus determine a range of \meff{} corresponding to 
\begin{equation}
(\meff_{\rm exp})_{\mbox{}_{\rm MIN}} - \Delta 
\le \meff  \le 
\zeta \Big( (\meff_{\rm exp})_{\mbox{}_{\rm MIN}} + \Delta \Big)~. 
\end{equation}
%
\noindent The currently estimated range of 
$\zeta^2$ for experimentally interesting 
nuclei varies from 3.5 for 
$^{48}$Ca to 38.7 for $^{130}$Te, 
see, e.g., Table 2 in \cite{ElliotVogel02}
and \cite{bilgri}. For $^{76}$Ge
and $^{82}$Se it is \cite{ElliotVogel02} 
$\sim 10$.

  In order to be possible to 
distinguish between the NH and IH spectra, 
between the NH and QD spectra, 
and between IH and QD spectra,
the following inequalities 
must hold, respectively:
\begin{equation} \label{eq:NHIHcond}
\zeta \, \meff_{\rm max}^{\rm NH} < \meff_{\rm min}^{\rm IH}~,
\end{equation}
\begin{equation} 
\zeta \, \meff_{\rm max}^{\rm NH} < \meff_{\rm min}^{\rm QD}~,
\end{equation}
\begin{equation} 
\zeta \, \meff_{\rm max}^{\rm IH} 
< \meff_{\rm min}^{\rm QD}~,~\zeta \ge 1~.
\label{eq:IHQDcond}
\end{equation}
%
\noindent These conditions imply, as it is 
not difficult to demonstrate 
\cite{PPRSNO2bb}, upper limits on $\ts$
which are functions of the neutrino 
oscillation parameters and of $\zeta$.  

    In Fig.\ \ref{fig:spr1} 
(taken from \cite{PPRSNO2bb}) 
the upper bounds on \ts{}, for which one 
can distinguish the NH
spectrum from the IH spectrum 
and from that of the QD type, 
are shown as a function of \dms$\!\!$
for $\dma = 3\times 10^{-3}~{\rm eV^2}$,
$\sin^2\theta = 0.05~{\rm and}~0.0$ and
different values of $\zeta$. 
For the NH vs IH spectrum 
results for  $\sin^2\theta = 0.01$
are also shown. In the case of the QD 
spectrum values of $m_0 = 0.2;~1.0$ eV 
are used. 

 As Fig.\ \ref{fig:spr1} 
demonstrates, the dependence of the 
maximal value of \ts{} of interest
on $m_0$ and $\sin^2\theta$ 
in the NH versus QD case is rather weak. 
This is not so in what concerns 
the dependence on $\sin^2\theta$
in the NH versus IH case: 
the maximal value of \ts{}
under discussion
can increase noticeably (e.g., by a 
factor of $\sim (1.2 - 1.5)$)
when $\sin^2\theta$ decreases from 0.05 to 0.
As it follows from 
Fig.\ \ref{fig:spr1}, it would 
be possible to distinguish between
the NH and QD spectra for the values of 
$\ts$ favored by the data
for values of $\zeta \cong 3$,
or even somewhat bigger than 3.
In contrast, the possibility to
distinguish between
the NH and IH spectra 
for $\zeta \cong 3$
depends critically on the value of
$\sin^2 \theta$: 
as Fig.  \ \ref{fig:spr1} indicates,
this would be possible
for the current best fit value
of $\ts$ and, e.g.,
$\deltasol = (5.0 - 15)\times 10^{-5}~{\rm eV^2}$,
provided $\sin^2\theta \ltap 0.01$.

 In Fig.\ \ref{fig:spr2} (taken from
\cite{PPRSNO2bb})
we show 
the maximal value of \ts{} 
permitting to distinguish between
the IH and QD spectra
as a function of \dma$\!\!$,
for $\sin^2\theta = 0.05~{\rm and}~ 0.0$,
$\deltasol = 7.0\times 10^{-5}~{\rm eV^2}$,
$m_0 = 0.2;~0.5;~1.0$ eV,
and $\zeta = 1.0;~1.5;~2.0;~3.0$.
The upper bound on $\ts$ 
of interest depends strongly 
on the value of
$m_0$. It decreases 
with the increasing
of \dma$\!\!$.
As it follows from Fig.\ \ref{fig:spr2},
for the values of 
\dma favored by the 
data and for $\zeta \gtap 2$, distinguishing
between the IH and QD spectra in the
case of $m_0 \cong 0.20$ eV requires 
too small, from the point of 
view of the existing data,
values  of $\ts$. For
$m_0 \gtap 0.40$ eV, the values of $\ts$ 
of interest fall in the ranges
favored by the  solar neutrino 
and KamLAND data
even for $\zeta = 3$.

  These quantitative analyses show 
that if \meff{} is found to be non-zero
in the future \betabeta-decay experiments,
it would be easier, in general, to distinguish 
between the spectrum with NH 
and those with IH or of QD type
using the data on $\meff \neq 0$,
than to distinguish between the 
IH and the QD spectra.
Discriminating between the 
latter would be less 
demanding if $m_0$ is sufficiently large.

\vspace{-0.4cm}
\section{Constraining the Majorana CP-Violation Phases}
\vspace{-0.3cm}

   The problem of detection of CP-violation 
in the lepton sector  
is one of the most formidable and challenging 
problems in the studies of neutrino mixing.
As was noticed in \cite{PPW}, 
the  measurement of \meff alone
could exclude the possibility of 
the two Majorana CP-violation phases 
$\alpha_{21}$ and $\alpha_{31}$,
present in the PMNS matrix
being equal to zero. However,
such a measurement cannot rule out
without additional input
the case of the two phases 
taking different 
{\it CP-conserving} values.
The additional input needed for
establishing CP-violation
could be, e.g., the measurement of 
neutrino mass $m_{\bar{\nu}_e}$ in 
\hbeta experiment
KATRIN \cite{MainzKATRIN}, or the 
cosmological determination of
the sum of the three neutrino masses \cite{Hu99},
$\Sigma = m_1 + m_2 + m_3$,
or a derivation of a sufficiently 
stringent upper limit on $\Sigma$.
At present no viable alternative to 
the measurement of 
\meff for getting information 
on the Majorana CP-violating phases 
$\alpha_{21}$ and $\alpha_{31}$
exists, or can be foreseen 
to exist in the next $\sim 8$  years.

   The possibility 
to get information 
on the CP-violation due to the Majorana
phases $\alpha_{21}$ and $\alpha_{31}$
by measuring \meff was studied 
by a large number of authors
\cite{BGKP96,BPP1,FViss00,WR00,bb0nuCP1},
and more recently, e.g., in  \cite{BargerSNO2bb,PPR1}.
The authors of \cite{BargerSNO2bb}
took into account in their analysis,
in particular, the effect of
the uncertainty in the knowledge of the
nuclear matrix elements 
on the measured value of \meff.
After making a certain
number of assumptions about the 
experimental and theoretical developments
in  the field of interest that may occur by
the year 2020
\footnote{It is supposed in \cite{BargerSNO2bb}, in particular, 
that \meff will be measured with a 25\% (1 s.d.) error and that
the uncertainty in the \betabeta-decay nuclear matrix elements 
will be reduced to a factor of 2.}, 
they claim to have shown ``once and for all
that it is impossible to detect CP-violation from 
\betabeta-decay in the foreseeable future.''
A different approach to the problem was
used in \cite{PPR1},
where an attempt was made to determine 
the conditions
under which CP-violation might be detected
from a measurement of \meff and 
$m_{\bar{\nu}_e}$ or $\Sigma$,
or of \meff and a sufficiently 
stringent upper limit $\Sigma$.
We will summarize the results 
obtained in the latter study.

 The analysis in \cite{PPR1} 
is based on
prospective input data on 
\meff$\!\!$, $m_{\bar{\nu}_e}$, $\Sigma$,
$\tan^2\theta_{\odot}$, etc.
The effect of the nuclear matrix element
uncertainty was included in the 
analysis.
For example, in the case of the inverted
hierarchical spectrum
($m_1 \ll m_2 \simeq m_3$, $m_1 < 0.02 \ \eV$),
a ``just-CP-violating'' region~\cite{BPP1} --- 
a value of \meff in this region would signal unambiguously
CP-violation in the lepton sector
due to Majorana CP-violating phases, would be present if
\begin{eqnarray}
(\meff_{\rm \!\!exp})_{\mbox{}_{\rm MAX}} <  \sqrt{\deltaatmmin}
\label{suffcondinv01a}
 \\
(\meff_{\rm \!\!exp})_{\mbox{}_{\rm MIN}} >  \sqrt{\deltaatmmax}
( \cos 2 \theta_{\odot} )_{\mbox{}_{\rm MAX}},
\label{suffcondinv01}
\end{eqnarray}
%
where $(\meff_{\rm \!\!exp})_{\mbox{}_{\rm MAX ( MIN)}}$
is the largest (smallest) experimentally 
allowed value of \meff$\!\!$,
taking into account both the experimental error
on the measured \betabeta-decay half life-time   
and the uncertainty due to
the evaluation of the nuclear matrix elements.
Condition (\ref{suffcondinv01})
depends crucially on the value of 
$( \cos 2 \theta_{\odot} )_{\mbox{}_{\rm MAX}}$
and it is less stringent for smaller values of 
 $( \cos 2 \theta_{\odot} )_{\mbox{}_{\rm MAX}}$
~\cite{PPW}. 

  Using the parametrization given in eq.~(\ref{eq:zeta}),
the necessary condition 
permitting to establish, in principle,
that the CP-symmetry is violated 
due to the Majorana CP-violating phases reads:
\begin{equation}
1 \leq \zeta <  \frac{\sqrt{\deltaatmmin}}
{\sqrt{\deltaatmmax}(\cos 2 \theta_{\odot})_{\mbox{}_{\rm MAX}} + 2 \Delta}~.
\label{invh:cond1a}
\end{equation}
Obviously, the smaller $(\cos 2 \theta_{\odot})_{\mbox{}_{\rm MAX}}$
and $\Delta$ the larger the ``theoretical uncertainty'' 
which might allow one to make conclusions 
concerning the CP-violation of interest.

A similar analysis was performed 
in the case of 
QD neutrinos mass spectrum.
The results can be summarized as follows.
 The possibility of 
establishing that the Majorana phases
$\alpha_{21}$ and $\alpha_{31}$ 
have CP-nonconserving values
requires quite accurate measurements
of \meff and, say, of $m_{\bar{\nu}_e}$ or $\Sigma$, 
and holds only for a limited range of 
values of the relevant parameters.
More specifically, 
proving that CP-violation associated with
Majorana neutrinos takes place
requires, in particular, a relative 
experimental error on the measured value of 
\meff not bigger than (15 -- 20)\%,
a ``theoretical uncertainty'' in the value of
\meff due to an imprecise knowledge of the 
corresponding nuclear matrix elements
smaller than a factor of 2, a value of 
$\tan^2\theta_{\odot} \gtap 0.55$,
and values of the relevant Majorana
CP-violating phases ($\alpha_{21}$, 
$\alpha_{32}$) typically 
within the ranges of $\sim (\pi/2 - 3\pi/4)$ and
$\sim (5\pi/4 - 3\pi/2)$. 

\vspace{-0.40cm}
\section{Conclusions} 
\vspace{-0.3cm}

 Future $\betabeta-$decay experiments 
have a remarkable physics potential. They can establish 
the Majorana nature of the neutrinos with definite mass
$\nu_j$. If the latter are Majorana particles,
the $\betabeta-$decay experiments can determine the type of the
neutrino mass spectrum and can provide unique
information on the absolute scale of neutrino masses.
They can also provide unique information on the
Majorana CP-violation phases present in the 
PMNS neutrino mixing matrix. The knowledge of the 
values of the relevant $\betabeta-$decay nuclear 
matrix elements with a sufficiently small uncertainty
is crucial for obtaining quantitative information on
the neutrino mass and mixing parameters 
from a measurement of
$\betabeta-$decay half life-time.
\vspace{-0.4cm}
\section{Acknowledgments}
\vspace{-0.3cm}
We are grateful to S. M. Bilenky, L. Wolfenstein and 
W. Rodejohann for very fruitful 
collaborations. S.T.P. would like to thank
M. Baldo Ceolin for organizing such a 
scientifically enjoyable Workshop.
This work was supported in part by 
the Italian MIUR and INFN 
under the programs ``Fenomenologia delle Interazioni Fondamentali'' 
and ``Fisica Astroparticellare'' (S.T.P.) and 
by the DOE Grant DE-FG03-91ER40662 (S.P.).

\begin{table}[ht]
\vspace{-0.6cm}
\begin{center}
\begin{tabular}{|c|c|c|c|c|c|c|} 
\hline
\rule{0pt}{0.5cm} Reference & $(\ts)_{\rm BF}$ & $(\dms\!\!)_{\rm BF}$  
& $\meff_{\rm max}^{\rm NH}$ & $\meff_{\rm min}^{\rm IH}$ & 
$\meff_{\rm min}^{\rm QD} $  
\\ \hline \hline
\protect\cite{fogli} & 0.46 & 7.3  
&  5.9 (3.9) & 18.4 & 59.9 \\ \hline
\protect\cite{band} & 0.42 & 7.2 
&  5.7 (3.7) & 20.3 & 67.2 \\ \hline
\protect\cite{bahcall} & 0.43 & 7.0 
&  5.7 (3.7) & 19.8 & 65.3 \\ \hline
\end{tabular}
\caption{\label{tab:BF} The best-fit values of $\ts$ and \dms 
(in units of $10^{-5} \; \rm eV^2$) 
in the LMA solution region,
as reported by different authors. Given are 
also the calculated maximal values of  
\meff (in units of $10^{-3}$ eV) for the NH spectrum 
and the minimal values of 
\meff (in units of $10^{-3}$ eV) 
for the IH and QD spectra.
The results for \meff in the cases of 
NH and IH spectra are obtained for 
$m_1 = 10^{-3}$ eV and the 
best-fit value of \dma$\!\!$,
$\dma = 2.7 \times 10^{-3} \eV^2$ 
\protect\cite{fogliold},
while those for 
the QD spectrum are derived for 
$m_0 = 0.2$ eV. 
In all cases $\sin^2\theta = 0.05$ 
has been used. For $\meff_{\rm max}^{\rm NH}$ we included in 
brackets also the values for $\sin^2\theta = 0.01$. 
The chosen value of $\dma$ corresponds to
$\meff_{\rm max}^{\rm IH} = 52.0 \times 10^{-3}$ eV.
(From \protect\cite{PPRSNO2bb}.)}
\end{center}
\end{table}

\begin{table}[ht]
\vspace{-0.7cm}
\begin{center}
\begin{tabular}{|c|c|c|c|c|c|c|} 
\hline
\rule{0pt}{0.5cm}Reference 
& $\ts$ & \dms  
& $\meff_{\rm max}^{\rm NH}$ & $\meff_{\rm min}^{\rm IH}$ & 
$\meff_{\rm min}^{\rm QD} $  
\\ \hline \hline
\cite{fogli} & 0.32 $-$ 0.72 & 5.6 $-$ 17   
&  8.6 (6.6) & 7.6 & 20.6 \\ \hline
\cite{band} & 0.31 $-$ 0.56 & 6.0 $-$ 8.7   
&  6.6 (4.5) & 13.0 & 43.2 \\ \hline
\cite{bahcall} & 0.31 $-$ 0.66 & 5.9 $-$ 8.9   
&  7.0 (4.9) &  9.5 & 28.6 \\ \hline
\end{tabular}
\caption{\label{tab:90} 
The ranges of allowed values of \ts and \dms 
(in units of $10^{-5} \; \rm eV^2$) 
in the LMA solution region, 
obtained at 90$\%$ C.L.\ by different authors.
Given are also the corresponding maximal values of  
\meff (in units of $10^{-3}$ eV)
for the NH spectrum, and the minimal values of 
\meff (in units of $10^{-3}$ eV) for the IH and QD spectra.
The results for the NH and IH spectra 
are obtained for $m_1 = 10^{-3}$ eV, while
those for the QD spectrum correspond to 
$m_0 = 0.2$ eV\@. 
~\dma was assumed
to lie in the interval \protect\cite{fogliold}
$(2.3 - 3.1) \times 10^{-3} \eV^2$. This implies 
$\meff_{\rm max}^{\rm IH} = 55.7 \times 10^{-3} \; \rm eV$.
As in Table 1, in all cases $\sin^2\theta = 0.05$ 
has been used. For $\meff_{\rm max}^{\rm NH}$ we included in 
brackets also the values for $\sin^2\theta = 0.01$.
(From \protect\cite{PPRSNO2bb}.)}
\end{center}
\end{table}

\begin{table}[ht]
\vspace{-0.95cm}
\begin{center}
\begin{tabular}{|c|c|c|c|c|c|} 
\hline
\rule{0pt}{0.5cm} Reference 
& $\meff_{\rm max}^{\rm NH} \, (s^2 = 0.05)$ 
& $\meff_{\rm max}^{\rm NH} \, (s^2 = 0.01)$ 
& $\meff_{\rm min}^{\rm IH}$ & 
$\meff_{\rm min}^{\rm QD} $  
\\ \hline \hline
\cite{fogli} 
& 6.1~(6.7) & 4.1~(4.4) & 16.5~(12.9) &  55.9~(48.2) \\ \hline
\cite{band}
& 6.0~(6.5) & 3.9~(4.2) & 18.3~(14.6) & 63.3~(55.9) \\ \hline
\cite{bahcall}
& 6.0~(6.5) & 3.9~(4.2) & 17.9~(14.1) & 61.4~(54.0)  \\ \hline
\end{tabular}
\caption{\label{tab:fut1} The values of 
$\meff_{\rm max}^{\rm NH}$, $\meff_{\rm min}^{\rm IH}$ and 
$\meff_{\rm min}^{\rm QD}$ 
(in units of $10^{-3}$ eV), calculated using   
the best-fit values of solar and atmospheric neutrino
oscillation parameters from Table \ref{tab:BF} 
and including 1 s.d.\ (3 s.d) 
uncertainties of 5 $\%$ (15\%)
on \ts and $\dms\!\!$, and
of 10 $\%$ (30\%) on \dma$\!\!$.
In this case one has: 
$\meff_{\rm max}^{\rm IH} = 54.5~(59.2) \times 10^{-3} \; \rm eV$.
(From \protect\cite{PPRSNO2bb}.)} 
\end{center}
\end{table}

\newpage
\clearpage
\begin{figure}
\begin{center}
\epsfig{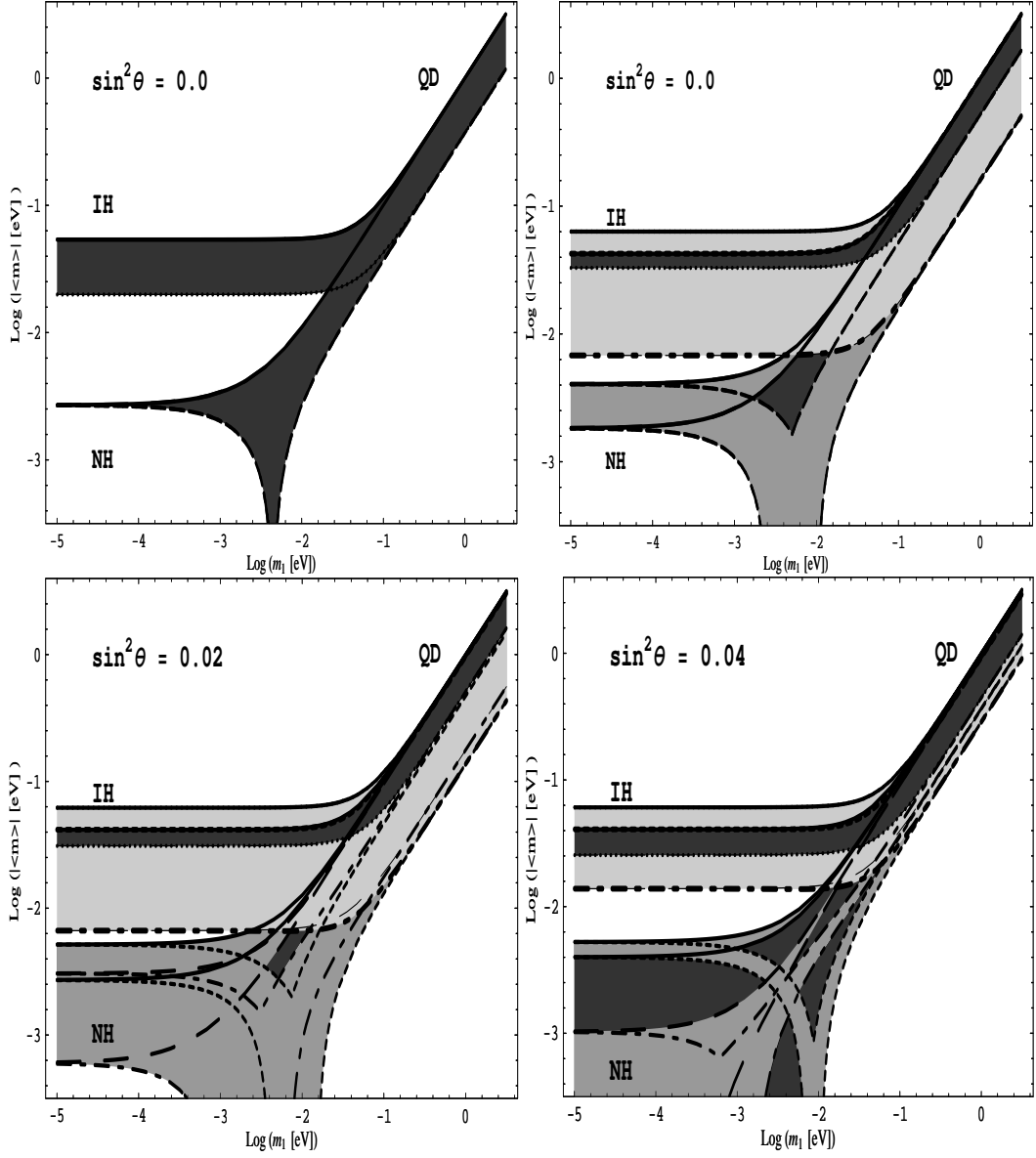}
\end{center}
\vspace{-4mm}
\caption{The dependence of \meff on $m_1$ 
for the solution LMA-I, 
$\deltasol = \Delta m_{21}^2$  and
$\deltasol = \Delta m_{32}^2$, and 
for the best fit
(upper left panel), and the $90 \%$~C.L. 
allowed (upper right and lower panels), 
values of the neutrino 
oscillation parameters found in 
refs. \protect\cite{fogli,fogliold}.
The values of $\sin^2 \theta$ used 
are $0.0$ (upper panels), $0.02$ (lower left panel)
and $0.04$ (lower right panel).
In the case of CP-conservation, \meff 
takes values: i) for the upper left panel
and $\deltasol = \Delta m_{21}^2$ 
($\deltasol = \Delta m_{32}^2$)
on {\it a)} the lower (upper) solid line 
 if $\eta_{21 (32)}=1$ and $ \eta_{31 (21)} = \pm 1$,
{\it b)} the long-dashed (dotted) line 
 if $\eta_{21 (32)}= - 1$ and $ \eta_{31 (21)} = \pm 1$;
ii) for the upper right panel
and $\deltasol = \Delta m_{21}^2$
($\deltasol = \Delta m_{32}^2$) -
in the medium grey (light grey) regions 
{\it a)} between the two lower solid lines
(the upper solid line and the short-dashed line)
if $\eta_{21 (32)}=1$ and $ \eta_{31 (21)} = \pm 1$,
{\it b)} between the two long-dashed lines
(the dotted and the dash-dotted lines)
 if $\eta_{21 (32)}= - 1$ and $ \eta_{31 (21)} = \pm 1$;
for the two lower panels
and $\deltasol = \Delta m_{21}^2$ - 
in the medium grey regions 
{\it a)} between the two lower solid lines
if $\eta_{21}= \eta_{31} = 1$,
{\it b)} between the long-dashed lines
if $\eta_{21}= -\eta_{31} = 1$,
{\it c)} between the two lower dash-dotted lines 
if $\eta_{21}= -\eta_{31} =- 1$,
{\it d)} between the two lower short-dashed lines 
if $\eta_{21}=  \eta_{31} = - 1$;
and iii) for the two lower panels
and $\deltasol = \Delta m_{32}^2$ - 
in the light grey regions delimited 
{\it a)} by the upper solid and the upper short-dashed
lines
if $\eta_{32}= \pm \eta_{21} = 1$,
{\it b)} by the dotted and the upper dash-dotted lines
if $\eta_{32}= \pm \eta_{21} = -1$.
Values of \meff in the dark grey regions 
signal CP-violation.
} 
\label{fig:1}
\end{figure}


\begin{figure}
\begin{center}
\epsfig{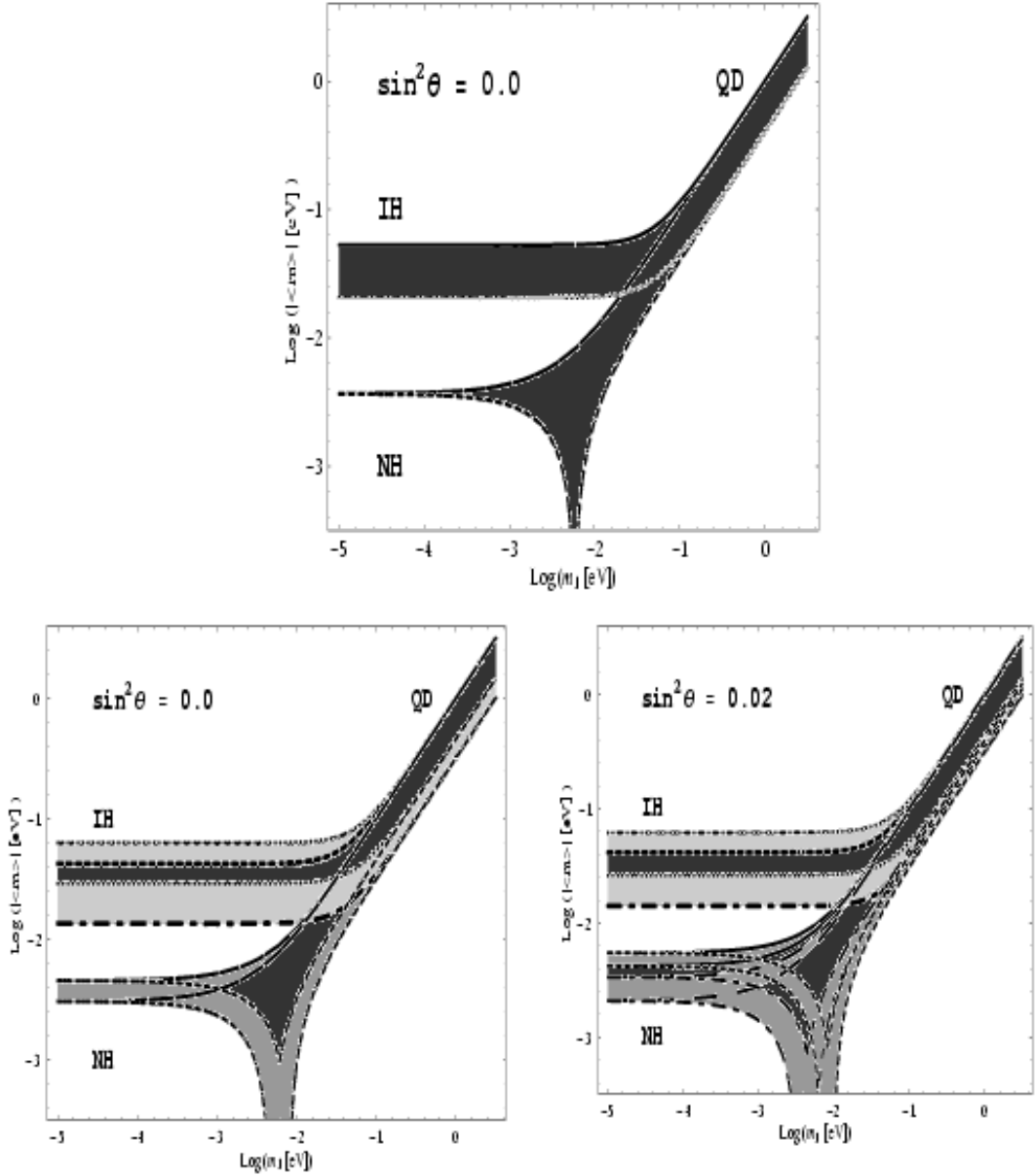}
\end{center}
\vspace{3mm}
\caption{The dependence of \meff on $m_1$ 
in the case of the solution LMA-II, 
for   $\deltasol = \Delta m_{21}^2$ and
$\deltasol = \Delta m_{32}^2$, and 
for the best fit values 
(upper panel) and the $90 \%$~C.L. allowed values 
(lower panels) of the neutrino 
oscillation parameters found in 
refs. \protect\cite{fogli,fogliold}.
The value of $\sin^2 \theta$ used 
are $0.0$ (upper and lower left panels)
and  $0.02$ (lower right panel).
In the case of CP-conservation,
the allowed values of \meff are constrained 
to lie on the same lines and regions 
as in Fig.~\ref{fig:1}:
for the upper (lower left) panel see the description
of the upper left (upper right) panel in Fig. 1,
and for the lower right panel refer 
to the explanations for the two lower 
panels in Fig. 1.  
Values of \meff in the dark grey regions 
signal CP-violation.
} 
\label{fig:2}
\end{figure}

\newpage
\clearpage

\begin{figure}
\begin{center}\hspace{-3cm}
\epsfig{file=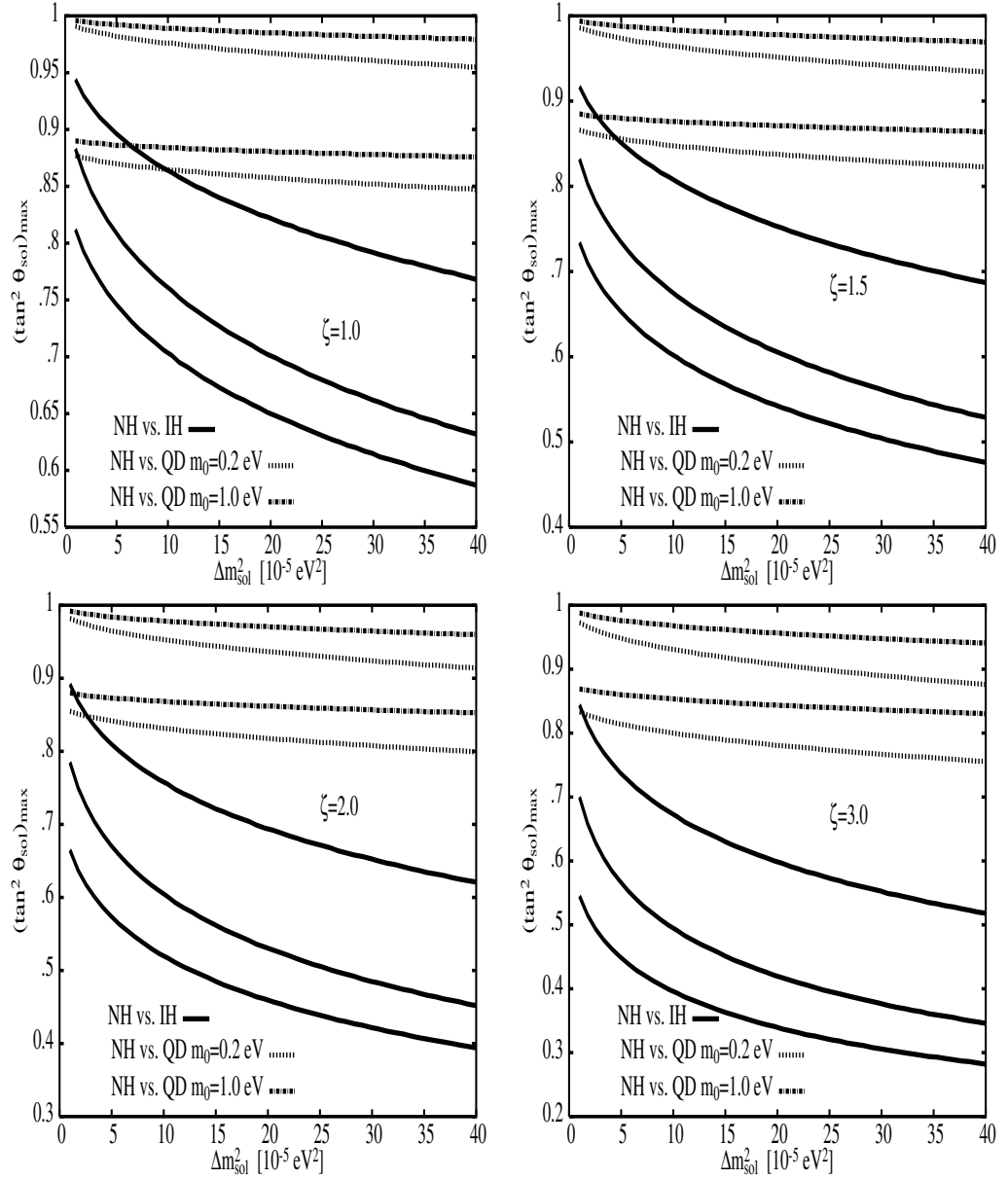,width=20cm,height=28cm}
\vspace{-10.5cm}
\caption{\label{fig:spr1} The upper bound on \ts{}, 
for which one 
can distinguish the NH
spectrum from the IH spectrum 
and from that of 
QD type, as a function of \dms$\!\!$
for $\dma\!\!= 3 \times 10^{-3} \eV^2$
and different values of $\zeta$.
The lower (upper) line corresponds to $\sin^2\theta = 0.05$ (0). 
For NH vs.\ IH there is a third (middle) 
line corresponding to  $\sin^2\theta = 0.01$.
(From \protect\cite{PPRSNO2bb}.) 
}
\end{center}
\end{figure}

\begin{figure}
\begin{center}\hspace{-2cm}
\epsfig{file=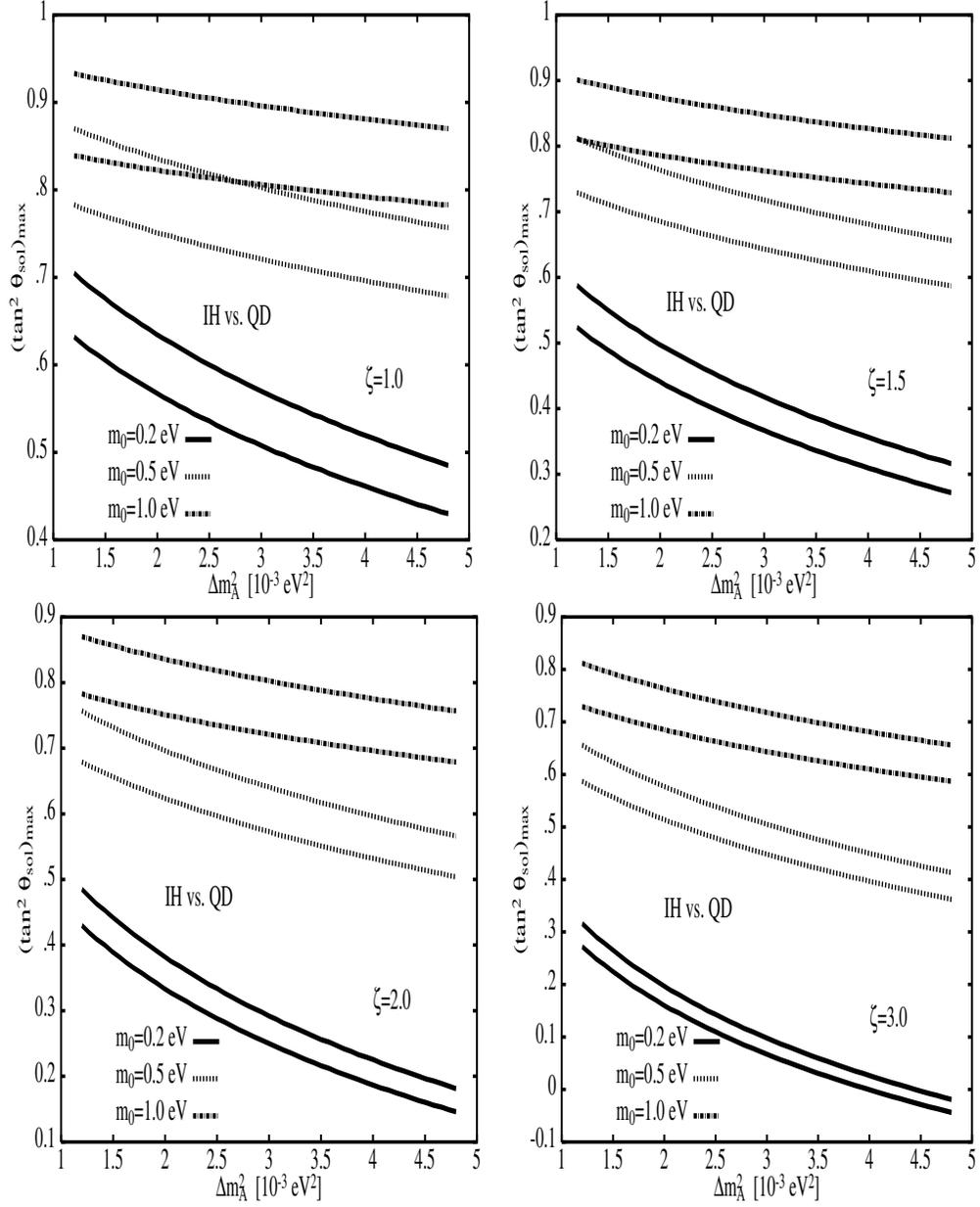,width=15cm,height=20cm}
\caption{\label{fig:spr2} The upper bound on \ts{} 
allowing one to discriminate between 
the IH and the QD neutrino mass spectra,
as a function of \dma for different 
values of $\zeta$. 
The lower (upper) line corresponds 
to $\sin^2\theta = 0.05$ (0).
(From \protect\cite{PPRSNO2bb}.)}
\end{center}
\end{figure}

\end{document}